\documentclass[12pt,aps,prd,tightenlines,groupedaddress,preprint,floatfix,nofootinbib]{revtex4}
\usepackage{amssymb,amsmath,graphicx,multirow,url}

\newcommand{\PRE}[1]{{#1}} 

\newcommand{\newc}{\newcommand}
\newc{\gsim}{\lower.7ex\hbox{$\;\stackrel{\textstyle>}{\sim}\;$}}
\newc{\lsim}{\lower.7ex\hbox{$\;\stackrel{\textstyle<}{\sim}\;$}}

\def\be{\begin{equation}}
\def\ee{\end{equation}}
\def\bea{\begin{eqnarray}}
\def\eea{\end{eqnarray}}

\def\IR{\relax{\rm I\kern-.18em R}}
 \font\cmss=cmss10 \font\cmsss=cmss10 at 7pt
\def\IQ{\relax{\rm I\kern-.18em Q}}
\def\IZ{\relax\ifmmode\mathchoice
 {\hbox{\cmss Z\kern-.4em Z}}{\hbox{\cmss Z\kern-.4em Z}}
 {\lower.9pt\hbox{\cmsss Z\kern-.4em Z}}
 {\lower1.2pt\hbox{\cmsss Z\kern-.4em Z}}\else{\cmss Z\kern-.4em Z}\fi}

\def\st1{\widetilde{t}_1}
\def\mst1{m_{\widetilde{t}_1}}
\def\Neut1{\widetilde{N}_1}
\def\N1{\widetilde{N}_1}

\newcommand{\gev}{{\rm GeV}}
\newcommand{\ev}{{\rm eV}}
\newcommand{\tev}{{\rm TeV}}
\newcommand{\kev}{{\rm keV}}

\newcommand{\pb}{{\rm pb}}
\newcommand{\m}{{\rm m}}

\newcommand{\fm}{{\rm fm}}
\newcommand{\s}{{\rm s}}
\newcommand{\km}{{\rm km}}

\newcommand{\sigmaSI}{\sigma_{\rm SI}}

\begin{document}

\preprint{UH-511-1192-12}

\title{\PRE{\vspace*{0.8in}}
Tools for Studying Low-Mass Dark Matter at Neutrino Detectors
\PRE{\vspace*{0.3in}}
}
\author{Jason Kumar}
\affiliation{Department of Physics and Astronomy, University of Hawaii, Honolulu, HI 96822 USA
\PRE{\vspace*{.5in}}
}

\author{John G.~Learned}
\affiliation{Department of Physics and Astronomy, University of Hawaii, Honolulu, HI 96822 USA
\PRE{\vspace*{.5in}}
}

\author{Katherine Richardson}
\affiliation{Department of Physics and Astronomy, University of New Mexico, Albuquerque, NM 87131 USA
\PRE{\vspace*{.5in}}
}

\author{Stefanie Smith}
\affiliation{Department of Physics and Astronomy, University of Hawaii, Honolulu, HI 96822 USA
\PRE{\vspace*{.5in}}
}

\begin{abstract}
\PRE{\vspace*{.3in}}
We determine the neutrino spectra arising from
low-mass ($4-10~\gev$) dark matter annihilating in the sun.
We also determine the low-mass dark matter capture rates (element by
element in the sun), assuming dark matter interacts either through
elastic contact interactions, elastic long-range interactions, or
inelastic contact interactions.  These are the non-detector-specific
data needed for determining the sensitivity of a neutrino detector to
dark matter annihilating in the sun.  As an application, we estimate the
sensitivity of a one kiloton liquid scintillation neutrino detector (such as KamLAND)
and LBNE (LAr-based) to low-mass dark matter with
long-range interactions and compare this to the expected CDMS sensitivity.  It is found that KamLAND's sensitivity can exceed that
obtainable from
the current CDMS data set by up to two orders of magnitude.
\end{abstract}


\maketitle



\input epsf



\section{Introduction}

There has been great recent interest in low-mass dark matter ($m_X \sim {\cal O}(10)~\gev$) as a
possible explanation for the event rates observed at the DAMA~\cite{Bernabei:2010mq},
CoGeNT~\cite{Aalseth:2011wp} and CRESST~\cite{Angloher:2011uu}
experiments.  However, negative results from XENON10/100~\cite{Angle:2011th,Aprile:2011hi}
and CDMS~\cite{arXiv:1010.4290,Ahmed:2010wy,Ahmed:2012vq} have created
some tension with these positive signals.  There has been much discussion in
the literature regarding experimental issues with all of this data, but the
explanation is far from
clear~\cite{Collar:2010gg,Collaboration:2010er,Collar:2010gd,Collar:2010ht,Collar:2011kf,Collar:2011wq,CollarTaup2011,Kuzniak:2012zm,Collar:2012ed}.
It is important to find new tests
of this data, particularly tests that do not suffer from the same difficulties as
direct detection experiments at low recoil energy.  Collider, gamma-ray, and anti-proton flux
search strategies have been employed for this
purpose~\cite{arXiv:0912.4510,Goodman:2010yf,arXiv:1007.5253,Goodman:2010ku,Rajaraman:2011wf,arXiv:1110.4376,Goodman:2011jq,Friedland:2011za,Kumar:2011dr,Shoemaker:2011vi}.

A promising avenue for cross-checking the low-mass direct detection data is with
neutrino detectors~\cite{Hooper:2008cf,Feng:2008qn,Fitzpatrick:2010em,Kumar:2011hi,Chen:2011vda,Agarwalla:2011yy},
which search for the flux of neutrinos arising from dark
matter annihilation in the core of the sun.  Dark matter is captured by
the sun via scattering off solar nuclei.  If the sun is in equilibrium,
then the dark matter capture rate determines the dark matter annihilation rate.  The
neutrino flux thus constrains the dark matter-nucleus scattering cross-section, and
allows neutrino detectors to cross-check direct detection experiments without
some of the particle physics and astrophysics uncertainties that plague other
types of indirect detection searches.  Moreover, the ${\cal O}(\gev)$ neutrinos
produced from the annihilation of low-mass dark matter are easily distinguishable
at water Cherenkov, liquid argon or liquid scintillator-based neutrino detectors.

In order to determine the event rate expected at a neutrino detector, one must
calculate the rate at which dark matter is captured by the sun due to scattering
and the neutrino spectrum arising from the decay of the dark matter annihilation
products.  Numerical packages such as DarkSUSY~\cite{Gondolo:2004sc} are commonly used for obtaining these
rates, which are determined from numerical simulations.  However, the required simulations
have not been run for masses less than 10 GeV.  Moreover, the capture rate has only
been calculated assuming that dark matter scatters elastically via a contact interaction.
Recent models for reconciling the low-mass direct detection data have included
the possibility of dark matter scattering via
long-range forces~\cite{Foot:2011pi,Fornengo:2011sz,Foot:2012rk},
inelastic scattering~\cite{TuckerSmith:2001hy,TuckerSmith:2004jv,Chang:2008gd,SchmidtHoberg:2009gn,Cline:2010kv,arXiv:1105.3734},
and dark matter interactions which are
isospin-violating~\cite{Fitzpatrick:2010em,Chang:2010yk,Feng:2011vu,arXiv:1105.3734,arXiv:1109.4639}.
To determine the sensitivity
of neutrino detectors to these models, new capture rate calculations must be performed.

In this work, we calculate the required neutrino spectra and capture rates for low-mass
dark matter in the sun.
We not only consider the capture rate for elastic contact scattering, but also for
inelastic dark matter and for models in which the dark matter-nucleon interaction
is mediated by a low-mass particle.
We also  determine the regions of parameter-space of these models for which dark matter in the sun
is in equilibrium.

As an application of these techniques, we consider the sensitivity of a 1 kT liquid scintillation
detector with 2135 live days of data (roughly the same exposure as KamLAND) to dark matter with
long-range interactions.  We find that the sensitivity of neutrino detectors to dark matter with
long-range interactions is enhanced, because typical scatters off low-mass targets (such as the
hydrogen and helium in the sun) involve small momentum transfers, yielding enhanced scattering
cross-sections in models where the mediating particles are light.  This implies the existence of an
entire class of dark matter models for which current neutrino experiments can provide the leading
sensitivity.

In section II, we review the general formalism for dark matter searches using neutrino detectors.  In
section III, we describe the details of the computation of the neutrino spectrum arising from the
annihilation of low-mass dark matter.  In section IV, we describe the calculation of the dark matter capture
rate in the sun, assuming dark matter with either elastic contact interactions, elastic long-range interactions,
or inelastic contact interactions.  In section V, we describe the range of circumstances under which low-mass dark
matter is in equilibrium in the sun.  In section VI, as an example, we apply these techniques to determine
the sensitivity of a 1 kT LS neutrino detector to low-mass dark matter with long-range interactions.
We conclude in section VII.

\section{Overview of Dark Matter Detection Via Neutrinos}

The rate of charged lepton events at a neutrino detector can be written as
\bea
R = {\Gamma_A \over 4\pi d^2} \int_0^1 dz\, \sum_{f,\nu_i} B_f {dN_{f,\nu_i} \over dz} \int dV\,
\sigma_{\nu_i -N}(z) \times \eta (r) \times \epsilon (r,z),
\eea
where $d \sim 1\,{\rm AU}$ is the earth-sun distance, $z= E_\nu / m_X$, $\eta$ is
the nucleon number density of the detector
(including the earth around the detector, in the case of a search for
throughgoing muons), and $\epsilon$ is the efficiency for a neutrino charged-current
interaction to produce a charged lepton which will pass the detector analysis
cuts.  $B_f$ is the branching
fraction to each dark matter annihilation product $f$, and $dN_{f,\nu_i} / dz$ is
the differential neutrino spectrum per annihilation to each final state, for each (anti-)neutrino
flavor.  $\Gamma_A$ is the total dark matter annihilation rate, and $\sigma_{\nu_i -N}$ is the
(anti-)neutrino-nucleon scattering cross-section.  For dark matter in the $4-10~\gev$ range, most
of the charged leptons produced in a reasonably-sized detector will be fully-contained, with the
vertex where the lepton is produced and the end of the track both within the
fiducial volume of the detector.  As a result, we will focus on fully-contained lepton events.

The quantities $\eta$ and $\epsilon$ are specific to the geometry and construction of the
detector.  For  $1~\gev < E_\nu <1~\tev$, the $\sigma_{\nu_i -N}$ can be approximated as~\cite{Edsjo:1993pb}
\bea
\sigma_{\nu -p} &\simeq& 4.51 \times 10^{-3}\times z\, (m_X /~\gev) ~\pb ,
\nonumber\\
\sigma_{\nu -n} &\simeq& 8.81 \times 10^{-3}\times z\, (m_X /~\gev) ~\pb ,
\nonumber\\
\sigma_{\bar \nu -p} &\simeq& 3.99 \times 10^{-3}\times z\, (m_X /~\gev) ~\pb ,
\nonumber\\
\sigma_{\bar \nu -n} &\simeq& 2.50 \times 10^{-3}\times z\, (m_X /~\gev) ~\pb .
\eea
The two remaining quantities we will need to compute are $\Gamma_A$ and $dN_f / dz$.

For a search for fully-contained charged leptons, we can then write
\bea
R = {\Gamma_A \over 4\pi d^2} \times  \int_0^1 dz\, \sum_{f,\nu_i} B_f {dN_{f,\nu_i} \over dz} z
\times A_{eff.}(z) ,
\eea
where
\bea
A_{eff.} (z) &=& \sigma_{\nu_i -N}(E_\nu = m_X) \int dV\,  \eta (r) \times \epsilon (r,z) .
\eea
All detector-specific information is encoded in the effective area, $A_{eff.}(z)$.

If $m_X \alt 4~\gev$, the effect of dark matter evaporation can be important~\cite{equilib,Hooper:2008cf}.
In that case, dark matter in the sun's core is significantly depleted by evaporation, and the total
annihilation rate is relatively small, implying that constraints on dark matter from the
neutrino flux will be weak.  We focus on the regime $m_X \geq 4~\gev$, for which dark matter
in the sun can only be depleted through annihilation.  If the sun is in equilibrium, we then
find that $\Gamma_A$ is related to the dark matter capture rate, $\Gamma_C$, by the relation
$\Gamma_C = 2\Gamma_A$.  Since $\Gamma_C$ is determined by the dark matter-nucleus scattering
cross-section, the above relation allows one to translate neutrino flux bounds into bounds on
the dark matter-nucleon scattering cross-section.

\section{Neutrino Spectra}

For low-mass dark matter ($m_X \leq 20~\gev$), the only dark matter annihilation
products relevant for neutrino searches are heavy quarks ($b$, $c$), $\tau$, $\nu_{e,\mu ,\tau}$
and the gluon ($g$).  Muons and light quarks ($u$, $d$ and $s$) will tend to stop within the
sun before they decay~\cite{spectrum}.  The resulting neutrinos are thus very soft (though they also 
can potentially be used for for dark matter searches~\cite{Rott:2012qb}).

We have calculated the differential neutrino spectrum per annihilation for the relevant channels
using the DarkSUSY/WimpSim/NuSigma/Pythia~\cite{Gondolo:2004sc,WimpSim,Sjostrand:2006za} package.
Numerical simulations were run on the Hawaii Open Supercomputing Center (HOSC) computing cluster.
$10^7$ annihilations were simulated for each annihilation channel and dark matter
mass in the range $4-10~\gev$ (in increments of 2 GeV).  Representative spectra are plotted in
Appendix A, and all of the original data files are available at
\url{http://www.phys.hawaii.edu/~superk/post/spectrum}.
We present the neutrino spectrum at a distance 1 AU from the sun, including the effects of
hadronization and decay of the annihilation products at injection, matter effects
as the neutrinos propagate through the sun (including tau-regeneration) and
vacuum oscillations.
The neutrino oscillation parameters were chosen to be
\bea
\theta_{12} &=& 33.2^\circ ,
\nonumber\\
\theta_{23} &=& 45^\circ ,
\nonumber\\
\Delta m_{21}^2 &=& 8.1 \times 10^{-5}~\ev^2 ,
\nonumber\\
\Delta m_{32}^2 &=& 2.2 \times 10^{-3}~\ev^2 ,
\nonumber\\
\theta_{13} &=& 10^\circ ,
\eea
assuming a normal hierarchy.
This choice is consistent with recent exciting data from the Daya Bay experiment~\cite{An:2012eh}
indicating $\theta_{13} \sim 9^\circ$.
Data files for the choice $\theta_{13}=0^\circ$ are also available online.
For this choice, the change in the neutrino spectrum is relatively small unless
dark matter annihilates directly to neutrinos in a flavor-dependent way.

In the case where dark matter annihilates to $b$-quarks, annihilation can only
proceed if the dark matter mass is larger than the mass of the $b$-hadron which
is produced.  Moreover, a $b$-quark will lose $\sim 27\%$ of its energy during
hadronization~\cite{spectrum,Jungman:1995df}.  Thus, the neutrino spectrum arising from annihilation
to $b$-quarks is simulated only for $m_X \geq 6~\gev$.
In the case where dark matter annihilates to $\tau \bar \tau$, the neutrino spectrum
has been computed by averaging over helicities.  Some dark matter candidates will
preferentially decay to certain helicities, which can have a significant effect on
the injected neutrino spectrum~\cite{Barger:2011em}.

The neutrino spectrum at 1 AU is equivalent to the spectrum of downward-going neutrinos
at the detector, averaged over the year.  This spectrum determines the
rate of downward going fully-contained charged leptons.  In the case where the charged lepton
is upward going through the earth, one should also include oscillation and matter effects as the
neutrinos pass through the earth.  These effects depend on the location of the detector and
can be determined by inputting the neutrino spectrum at 1 AU into the ``WimpEvent" program,
with the location of the detector specified.

\section{Capture Rate}

The capture rate can be computed following the analysis of~\cite{Gould:1987ir}, and we
follow that notation.  A dark matter particle in the halo has velocity $u$, given by a
distribution $f(u)$ obeying $\int du\, f(u) = \eta_X$.  Here, $\eta_X$ is the dark
matter number density in the halo.  When a dark matter particle is at distance $r$ from
the core of the sun, it will have velocity $w=\sqrt{u^2 +v^2}$, where $v(r)$ is the escape
velocity from the sun at radius $r$.  Thus, a dark matter-nucleus scatter will result in
dark matter capture if the dark matter scatters from velocity $w$ to a velocity $\leq v$.

More generally, however, 3-body interactions can drastically affect the capture rate.
As a conservative estimate, one can choose to count as ``captured" only dark matter
particles which are kinematically constrained to orbits with maximum radius $r_0$, often
taken to be the radius of Jupiter's orbit.  The velocity to escape from position $r<r_0$ within the sun
to radius $r_0$ is denoted by $v_e (r)$, and is given by the relation
$v_e (r)^2 = v(r)^2 - v(r_0)^2$.

The rate for dark matter to be captured in any differential solar volume by scattering off
an element with atomic number $Z$ can then be written
as
\bea
{dC_Z \over dV} &=& \int_{u_{min}}^{u_{max}} du \, {f(u) \over u} w \Omega_v^- (w) ,
\eea
where $u_{min,max}$ are the minimum/maximum dark matter velocities in the halo such that a nuclear
scatter at position $r$ resulting in dark matter capture is kinematically possible.
$\Omega_v^- (w)$ is the rate per unit time at which a dark
matter particle with velocity $w$ will scatter to velocity $<v_e (r)$,
and is given by the expression
\bea
\Omega_v^- (w) &=& \eta_\odot w \int_{E_{min}}^{E_{max}} dE_R {d\sigma^{Z,A} \over dE_R},
\eea
where $\eta_\odot$ is the number density of the sun, $E_{min}$ is the minimum
recoil energy needed for capture, and $E_{max}$ is the maximum recoil energy
that is kinematically allowed.
$d\sigma^{Z,A} / dE_R$ is the differential cross-section for dark matter to
scatter off a nucleus with $Z$ protons and $A$ nucleons.

For any dark matter model, the quantities that
must be known to compute the capture rate are ${d\sigma^{Z,A} / dE_R}$, $E_{min,max}$,
and $u_{min,max}$.  Given these quantities, we can compute the capture rate numerically
using the DarkSUSY code (appropriately modified), with its standard assumptions about
solar composition.

\subsection{Elastic Contact Interactions}

The most commonly used assumption is that dark matter interacts with nuclei via
elastic, isospin-invariant, contact interactions.  In this case, we have
\bea
E_{min} &=& {1\over 2} m_X \left( u^2 + v(r_0)^2 \right) ,
\nonumber\\
E_{max} &=&  {\mu \over \mu_+^2} \left({1\over 2} m_X w^2 \right) ,
\nonumber\\
{d\sigma^{Z,A} \over dE_R} &=& {\sigma^p \over E_{max}}
\left[{(m_X + m_p)^2 \over (m_X +m_A)^2}{m_A^2 \over m_p^2}\right]\left[Z + {f_n \over f_p}(A-Z)\right]^2
|F_A(E_R)|^2 ,
\nonumber\\
u_{max}^2 &=&  {\mu \over \mu_-^2} \left(v^2 -{\mu_+^2 \over \mu }v(r_0)^2\right) ,
\nonumber\\
u_{min} &=& 0 ,
\label{ElasticContactEqns}
\eea
where $\mu \equiv {m_X / m_A}$ and $\mu_\pm \equiv {(\mu \pm 1) / 2}$.
$f_{p,n}$ are the relative strengths of dark matter coupling to protons and
neutrons, respectively.
$\sigma^p$ is the
dark matter-proton scattering cross-section, and $F_A(E_R)$ is the
nuclear form factor.
To match the assumptions used in DarkSUSY, we will assume a Gaussian form factor
$|F_A(E_R)|^2 = \exp [-E_R / E_0]$, where
$E_0 = 3 \hbar^2 / 2 m_A R_A^2$.
$R_A = [0.3 +0.91 (m_A /~\gev)^{1/3}]~\fm$ is taken as the nuclear radius.

Following default assumptions in
DarkSUSY~\cite{Gould:1987ir,Gondolo:2004sc},
we have assumed a Maxwell-Boltzmann velocity distribution for dark matter in the Galactic halo
of the form
\bea
f_{halo}(u) &=& \eta_X {4 \over \sqrt{\pi}} \left({3\over 2} \right)^{3\over 2} {u^2 \over \bar v^3}
e^{-3u^2 /2\bar v^2} ,
\eea
where $\bar v$ is the three-dimensional velocity dispersion, which we have set to
$\bar v = 270~\km/\s$.
This velocity
distribution is truncated at the galactic escape velocity, $v_{esc}$.
The velocity distribution seen by an observer moving through the halo with velocity $v_*$ is then
\bea
f(u)
&=& {f_{halo}(u) \over 2}  e^{-3v_*^2 /2\bar v^2} {\bar v^2 \over 3 u v_*}
\left[\exp\left({3u v_*  \over\bar v^2} \cos \theta_{max} \right)-
\exp\left({3u v_*  \over\bar v^2} \cos \theta_{min} \right) \right]
\nonumber\\
&\,& \times \theta (v_* + v_{esc} -u) ,
\eea
where
\bea
\cos \theta_{max} &=& 1 ,
\nonumber\\
\cos \theta_{min} &=& \max \left[-1, {u^2 - (v_{esc}^2 - v_*^2)
\over 2 u v_*} \right].
\eea
The step function imposes the condition that $f(u)=0$ for $u > v_* + v_{esc}$.
If we ignore the truncation at the galactic escape velocity, this
reduces to the expression
\bea
f(u) &=& f_{halo}(u) e^{-3v_*^2 /2\bar v^2} {\bar v^2 \over 3uv_* } \sinh \left({3uv_* \over \bar v^2} \right) .
\eea
We take $v_* = 220~\km/\s$ to be the velocity of the sun through the halo.

For isospin-invariant interactions, one assumes $f_n / f_p =1$.  For isospin-violating
dark matter, the capture rate from scattering by each element is scaled by a factor
$[Z+(f_n / f_p)(A-Z)]^2/A^2$.  To facilitate this rescaling in the case of generic
isospin-violating interactions, we plot the capture rate for each of the main elements
in the sun separately.

\begin{figure}[tb]
\includegraphics[width=0.95\columnwidth]{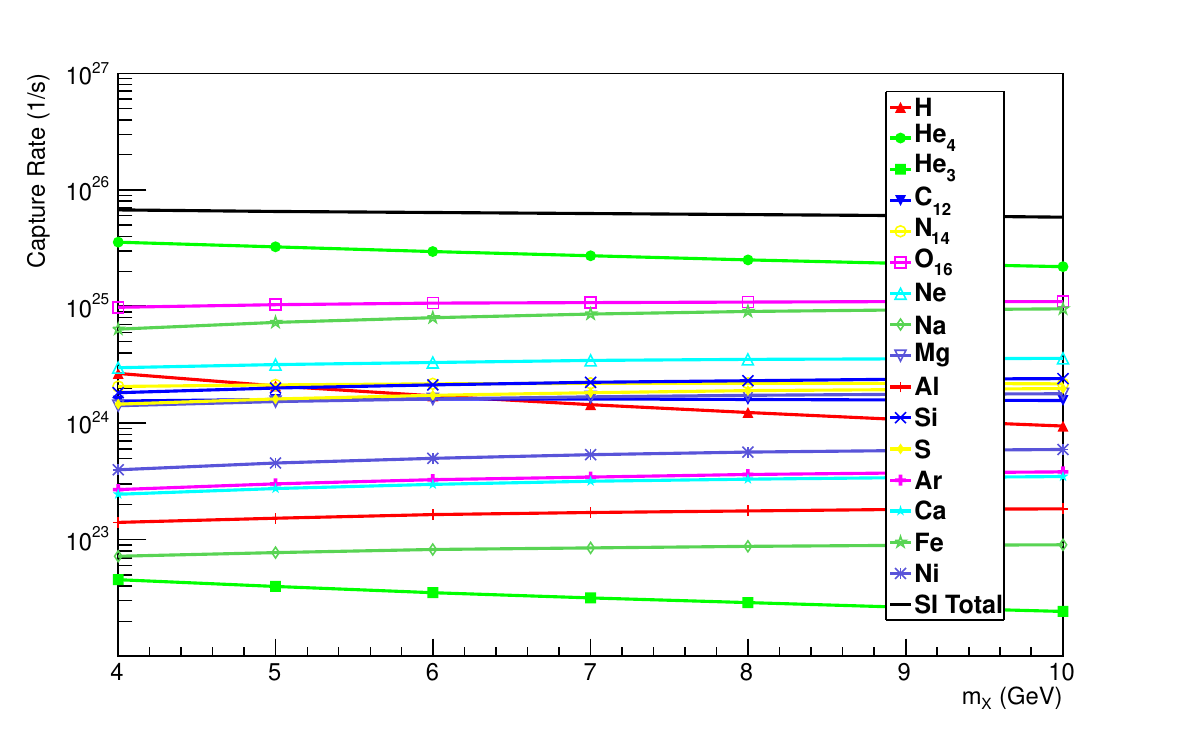}
\vspace*{-.1in}
\caption{Solar dark matter capture rates for various elements in the sun, assuming
isospin-invariant elastic contact interactions with $\sigma^p = 10^{-4}~\pb$.}
\label{CapPlotContact}
\end{figure}

We have plotted in fig.~\ref{CapPlotContact} the capture rates for elastic contact interactions if one requires
dark matter to be captured to within the radius of Jupiter's orbit.  If the presence of Jupiter is neglected, capture rates
change by less than $1\%$ in the case of elastic contact interactions.

\subsection{Elastic Long-Range Interactions}

\begin{figure}[tb]
\includegraphics[width=0.95\columnwidth]{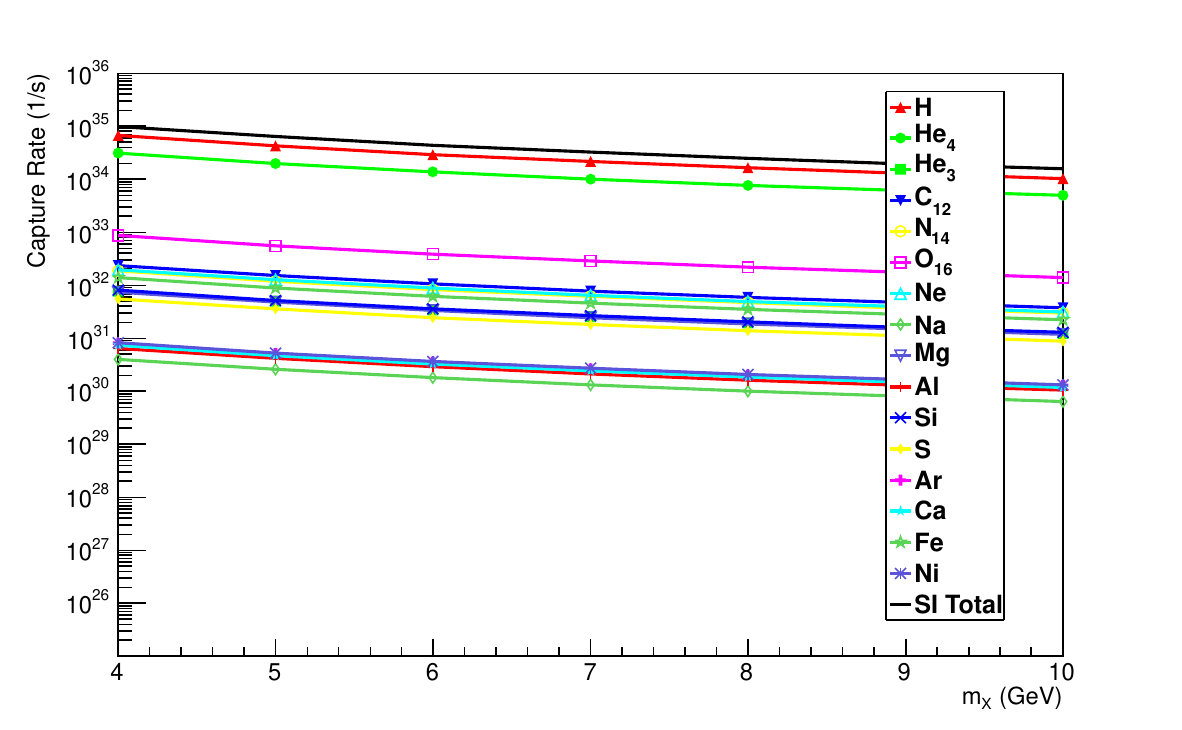}
\vspace*{-.1in}
\caption{Solar dark matter capture rates for various elements in the sun, assuming
isospin-invariant elastic long-range interactions (as in eq.~\ref{LongRangeScatteringEq}) with $C=10^{-10}$.
Captured dark matter is required to be confined within the orbit of Jupiter.}
\label{CapPlotLongRange}
\end{figure}

\begin{figure}[tb]
\includegraphics[width=0.95\columnwidth]{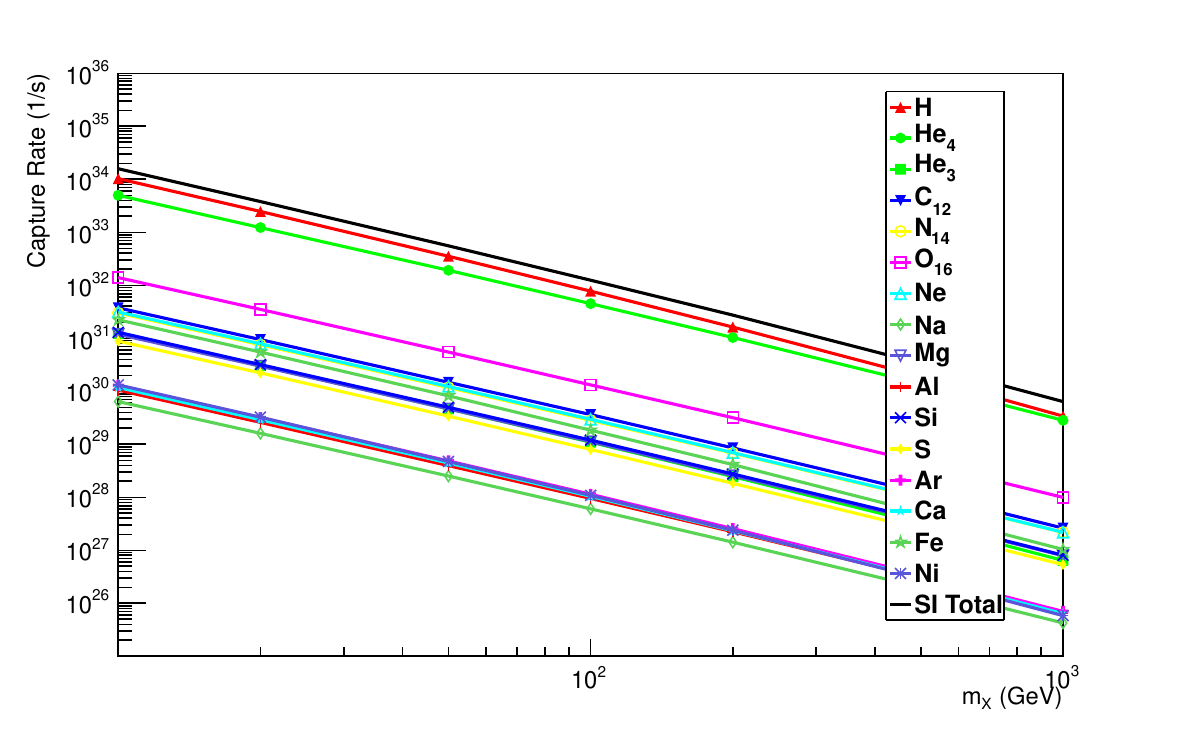}
\vspace*{-.1in}
\caption{Solar dark matter capture rates for various elements in the sun, assuming
isospin-invariant elastic long-range interactions (as in eq.~\ref{LongRangeScatteringEq}) with $C=10^{-10}$.
Captured dark matter is required to be confined within the orbit of Jupiter.
Plotted for dark matter with mass in the range $10-1000~\gev$.}
\label{CapPlotLongRange1000}
\end{figure}

If dark matter interacts with nuclei via long-range interactions (or, equivalently,
via a $t$-channel interaction with a mediator with a mass much smaller than
the momentum transfer), then the
differential scattering cross-section will have a different form.
The quantum matrix element scales as ${\cal M} \propto 1/(q^2 -M_*^2)$ where $M_*$ is the
mass of the mediating particle and $q$ is the momentum transfer; for $M_*^2 \ll q^2$, the
differential cross-section scales as $q^{-4}$.
For such a model, it
would not make sense to parameterize the differential scattering cross-section in terms of
$\sigma_p$ since, as with Rutherford scattering, the total cross-section is infinite.
We may instead write
\bea
{d\sigma^{Z,A} \over dE_R} &=& C {4\pi \alpha^2 \mu_p^2 \over m_A^2 E_R^2 E_{max}}
\left[{(m_X + m_p)^2 \over (m_X +m_A)^2}{m_A^2 \over m_p^2}\right]\left[Z + {f_n \over f_p}(A-Z)\right]^2
|F_A(E_R)|^2 ,
\label{LongRangeScatteringEq}
\eea
where $\mu_p = m_X m_p / (m_X + m_p)$ is the dark matter-proton reduced mass.
$C$ is a constant which defines the size of the differential scattering cross-section in terms
of the proton charge; if $g_{X,p}$ are the
strengths with which the mediator couples to the dark matter and a proton, respectively, then
$C = g_X^2 g_p^2 / e^4$.
Using the Gaussian form factor defined above, the integrated differential scattering cross-section takes the
simple form
\bea
\int_{E_{min}}^{E_{max}} dE_R \, {d\sigma^{Z,A} \over dE_R} &=& C{4 \pi \alpha^2 \mu_p^2 \over m_A^2 E_{max} }
\left[{(m_X + m_p)^2 \over (m_X +m_A)^2}{m_A^2 \over m_p^2}\right]\left[Z + {f_n \over f_p}(A-Z)\right]^2
\nonumber\\
&\,& \times
\left[{e^{-{E_{min}\over E_0}} \over E_{min}} - {e^{-{E_{max}\over E_0}} \over E_{max}} +{Ei(-E_{min}/E_0) \over E_0}
-{Ei(-E_{max}/E_0) \over E_0} \right] .
\eea
The kinematics of the scattering process
are the same as in the case of an elastic contact interaction, and thus $E_{min,max}$ and
$u_{min,max}$ are the same as in equation (\ref{ElasticContactEqns}).  The capture rates for dark matter with long-range
interactions are plotted in fig.~\ref{CapPlotLongRange} (for $m_X = 4-10~\gev$) and fig.~\ref{CapPlotLongRange1000}
(for $m_X = 10-1000~\gev$), again assuming that captured dark matter must be confined
to an orbit inside Jupiter's.

One should note that, in the case of long-range interactions, it is necessary to assume that
captured dark matter be confined to an orbit within some finite radius $r_0$; without this assumption,
the capture rate would be infinite.  The origin of this divergence is easily understood to arise from the low-velocity
tail of the Maxwell-Boltzmann velocity distribution.  Near $u=0$, dark matter far from the sun has a very small
kinetic energy.  As a result, even scattering interactions yielding very small recoil energies can result
in a dark matter particle being captured, (i.~e., having negative total energy).  Since the differential scattering
cross-section diverges at small recoil energy, the total capture rate diverges.  This simply reflects the fact
that it is not physically sensible to think of dark matter as captured if confined to an orbit of very
large radius.  It is most sensible to count as captured only dark matter confined to orbits that lie within
Jupiter's orbit.

Note that we are not including the possibility of capture due to multiple scattering.  For many models,
these effects can significantly enhance the dark matter capture rate, especially in the case of long-range
interactions.  The dark matter capture rate may be much larger if dark matter scattering exhibits Sommerfeld
enhancement.  But this depends on the details of the model, including the nature of dark matter interactions
with electrons and possible 3-body effects.  These issues may be relevant for specific models but are
beyond the scope of this work.

\subsection{Inelastic Contact Interactions}

One may also consider the case where dark matter scatters inelastically off nuclei, via the
process $X A \rightarrow X' A$.
We will consider the case with $\delta m_X = m_{X'} - m_X \geq 0$.
In this case, the scattering matrix element will only change
by subleading $ {\cal O}( \delta m_X / m_X) $ terms, but the kinematics of the scattering process can change dramatically.
It is easiest to consider this process in the center-of-mass frame.  We then find
$p_i^2 - p_f^2 \approx 2m_r \delta m_X$, where $m_r = m_X m_A / (m_X + m_A)$ is the
reduced mass, and $p_i =m_r w$ and $p_f$ are the spatial momenta of the incoming $X$ and outgoing $X'$,
respectively, in the center-of-mass frame.  The phase space factor of the differential scattering
cross-section is directly proportional to the outgoing momenta.

Again, it is not appropriate to express the dark matter-nucleus inelastic scattering cross-section in terms
of the dark matter-proton scattering cross-section, since there exist kinematic regions where
dark matter-proton inelastic scattering is impossible, though dark matter can scatter off other
nuclei.  Using the fact that the recoil energy can be written
as $E_R =(2m_A)^{-1} (p_i^2 + p_f^2 - 2p_i p_f \cos \theta_{cm})$, we can write
\bea
{d\sigma^{Z,A} \over dE_R} &=& {m_A I\over 32\pi w^2}
\left[Z + {f_n \over f_p}(A-Z)\right]^2
|F_A (E_R)|^2 ,
\eea
where $m_X^2 m_A^2 [Z + (A-Z)(f_n/ f_p)]^2 |F_A (E_R)|^2 \times I$ is the squared dark matter-nucleus matrix element
(summed over final spins and averaged over initial spins).  $I$ is
roughly constant for different elements (up to ${\cal O}(\delta m_X / m_X)$ corrections), so
it makes sense to present bounds on inelastic dark matter
in terms of this quantity.
We also find
\bea
E_{max} &=& {m_{r}^2 w^2\over m_A } \left(1 - {\delta m_X \over m_{r} w^2}
+  \sqrt{1- 2 {\delta m_X \over m_{r} w^2}}  \right) .
\eea
Similarly, we find
\bea
E_{min} &=& \max \left[{m_{r}^2 w^2\over m_A } \left(1 - {\delta m_X \over m_{red.} w^2}
-  \sqrt{1- 2 {\delta m_X \over m_{red.} w^2}}  \right), {1\over 2}m_X \left(u^2 +v(r_0)^2 \right)- \delta m_X \right]
\eea
where the first term is the minimum recoil energy kinematically possible in two-body inelastic scattering,
and the second term is the minimum recoil energy in a process where the outgoing dark matter
particle is slower than $v_e$, the velocity to escape to radius $r_0$.

$u_{max}$ is determined by the constraint $E_{max} \geq E_{min}$, yielding
\bea
u_{max}^2 &=& {1\over 2} v^2 {\mu \over \mu_-^2}
\left[ 1- {2\delta m_X \over m_X v^2 }\right.  {\mu_- \over \mu}
-{(1 + \mu^2) v(r_0)^2 \over 2\mu v^2}
\nonumber\\
&\,&
+ \left. \sqrt{\left(1-{v(r_0)^2 \over v^2}\right)^2    -4 {\delta m_X \over  m_X} {\mu_- \over v^2 }
\left( 1 - {v(r_0)^2 \over v^2} \right)     } \right] .
\eea
Finally, we have
\bea
u_{min}^2 = \max \left[ {2\delta m_X \over m_r } -v^2 ,0 \right],
\eea
because inelastic scattering is only kinematically possible for $\delta m_X \leq (1/2)m_r w^2$.
This implies that, for $m_X \alt 10~\gev$ one need only consider models with $\delta m_X \alt {\cal O}(10-100)~\kev$.

\begin{figure}[tb]
\includegraphics*[width=1.00\columnwidth]{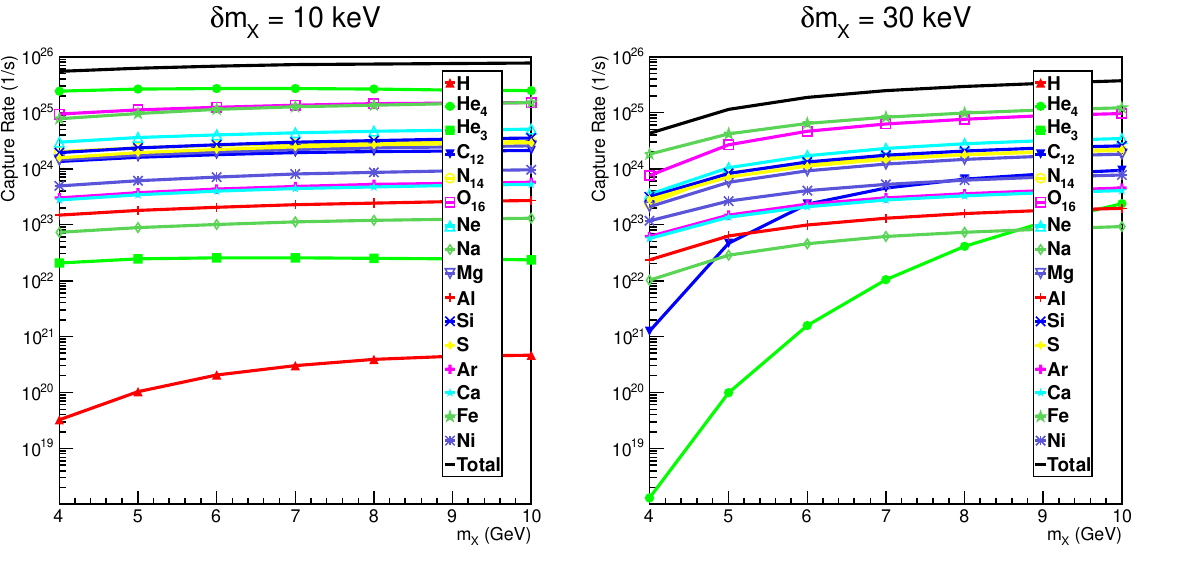}
\includegraphics*[width=0.50\columnwidth]{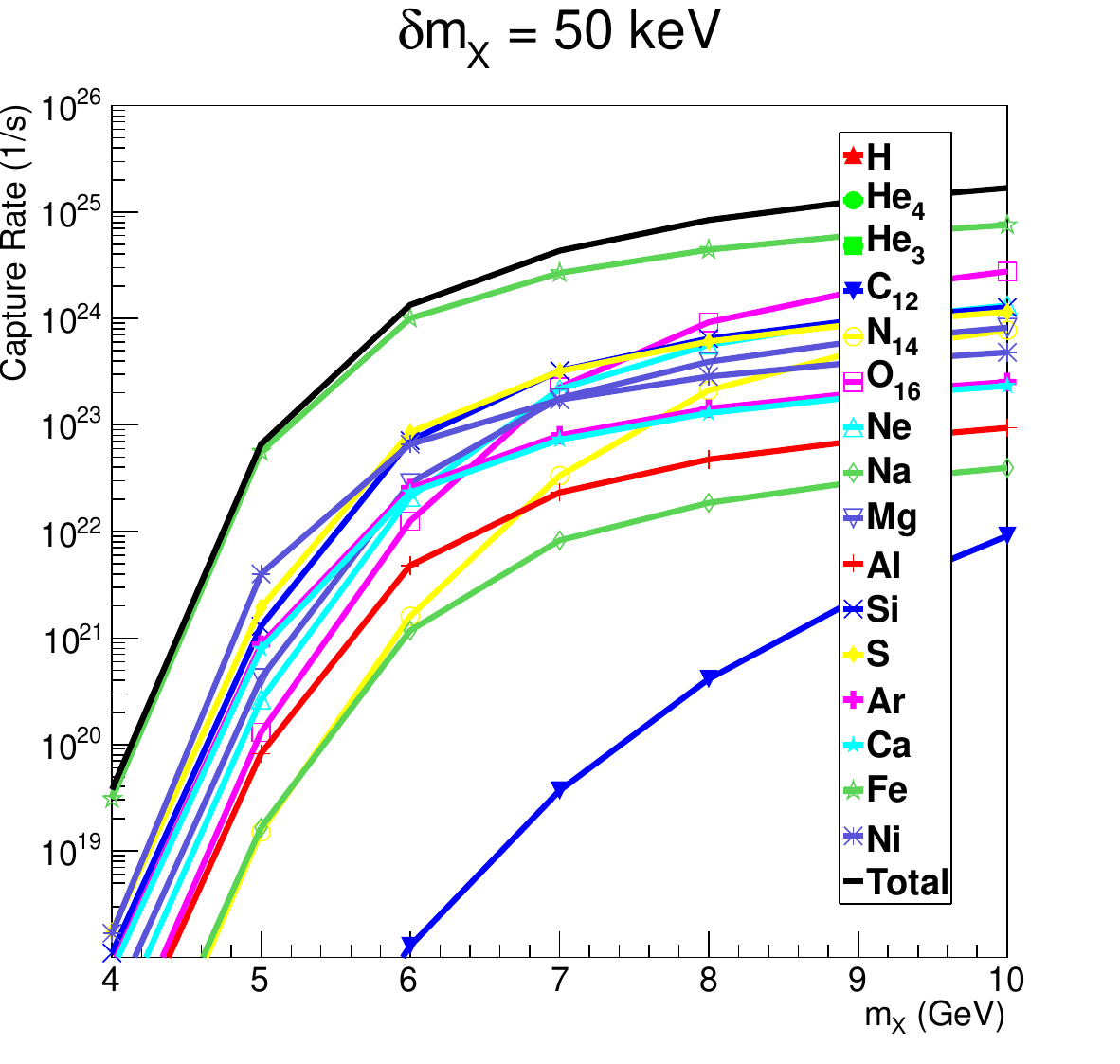}
\vspace*{-.1in}
\caption{Solar dark matter capture rates for various elements in the sun, assuming
isospin-invariant inelastic elastic contact interactions with $I / 32\pi = 10^{-4}~\pb ~\gev^{-2}$.
The three panels are for $\delta m_X = 10, 30,$ and $50~\kev$, as labelled.}
\label{CapPlotInelastic}
\end{figure}

The capture rates for inelastic dark matter with contact interactions and
$\delta m_X = 10,~30,~50~\kev$ are plotted in fig~\ref{CapPlotInelastic}.
It is interesting to note that, as $\delta m_X$ increases, the rate of capture arising from
scattering off light elements vanishes.  This is because $m_r$ is smallest for light elements, implying
that they have the smallest maximum value of $\delta m_X$ such that inelastic scattering is kinematically
allowed.  It is also worth noting that inelastic scattering for low-mass dark matter is kinematically
allowed in the sun for larger $\delta m_X$ than in the Earth, because dark matter within the sun
has gained kinetic energy from gravitational infall.  As a result, even low-mass inelastic dark matter with
$\delta m_X \sim 50~\kev$ can be potentially probed by neutrino detectors.

\section{Equilibrium}

If the effects of WIMP evaporation are negligible, the equilibration time $\tau_\odot$ for
dark matter in the sun can be written as~\cite{Griest:1986yu,Jungman:1995df}
\bea
{t_\odot \over \tau_\odot} &=& 1.9 \times 10^{-11} \left( {\Gamma_C^\odot \over {\rm s}^{-1}} \right)^{1\over 2}
\left({\langle \sigma v \rangle \over \pb } \right)^{1\over 2}
\left(m_X \over 10~\gev \right)^{3\over 4} ,
\eea
where $t_\odot \sim 4.5\times 10^9~{\rm yr}$ is the age of the solar system.

\begin{figure}[tb]
\includegraphics[width=0.75\columnwidth]{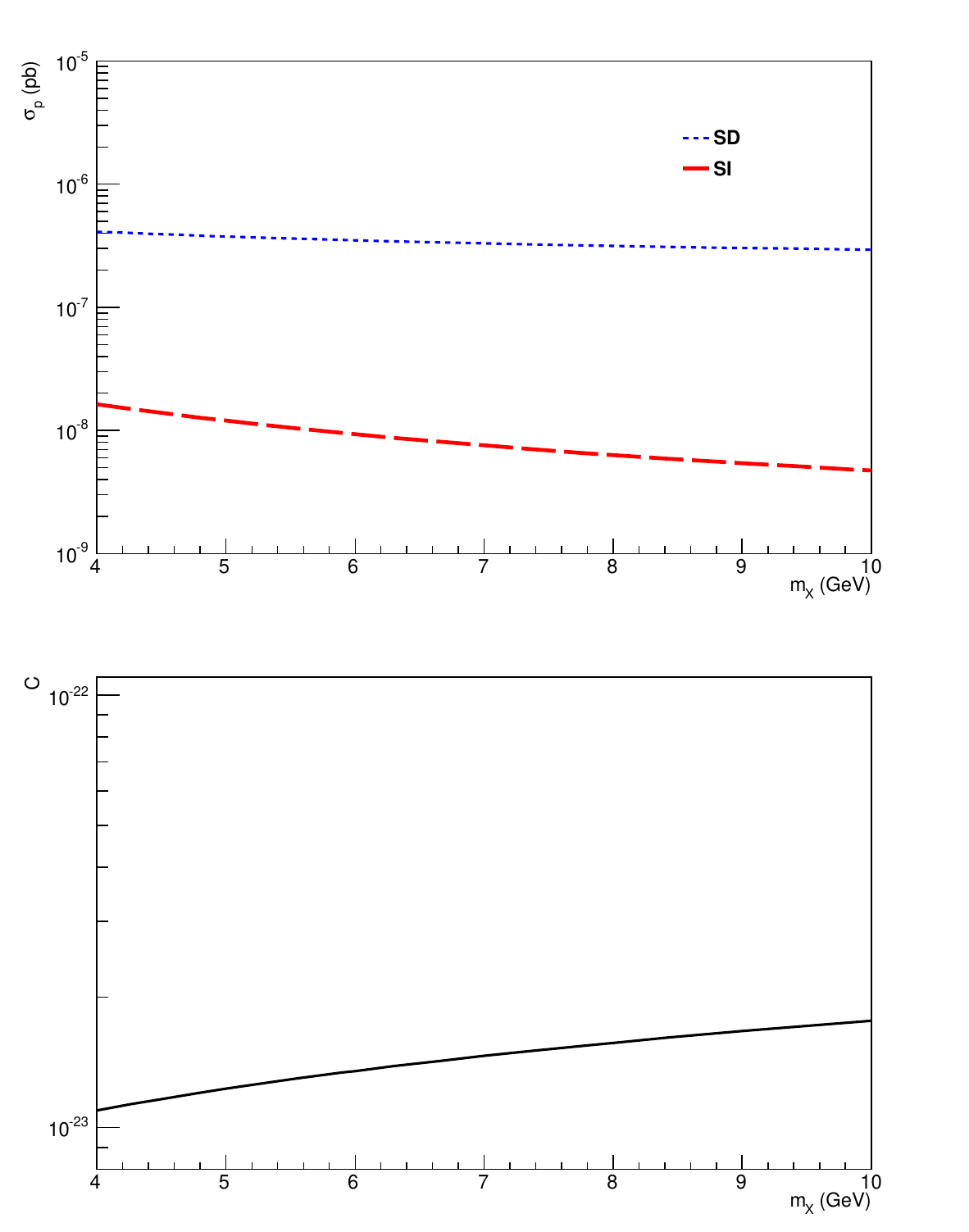}
\vspace*{-.1in}
\caption{Minimum $\sigma^p$ (top panel) or $C$ (bottom panel) required for dark matter to be in equilibrium
in the sun, assuming either elastic contact (spin-dependent or spin-independent) or
long-range interactions, respectively.  We assume that the dark matter annihilation cross-section is
given by $\langle \sigma v \rangle = 1~\pb$.}
\label{Equilibrium}
\end{figure}

For the case of elastic contact interactions (assumed to be  isospin-invariant, and either
spin-independent or spin-dependent), we plot in the top panel of
fig.~\ref{Equilibrium} the minimum $\sigma^p$ required for the sun to currently be in equilibrium, assuming that the
total dark matter annihilation cross-section is given by $\langle \sigma v \rangle = 1~\pb$.  Note that,
for the case of IVDM with spin-independent interactions, the $\sigma^p$ required for the sun to be
in equilibrium would lie between that required for spin-dependent scattering and that required for isospin-invariant
spin-independent scattering.
Note that the equilibration time scales as $(\sigma^p \langle \sigma v \rangle)^{-1/2}$.
IVDM with spin-independent interactions and $f_n / f_p \sim -0.7$ could be consistent with the data of
DAMA, CoGeNT and XENON10/100 if $\sigmaSI^p \sim 10^{-2}~\pb$~\cite{Chang:2010yk,Feng:2011vu}.
Such dark matter can be in equilibrium in the sun even if
$\langle \sigma v \rangle \sim 10^{-5}~\pb$.  This implies that, if the IVDM candidate is a thermal
relic, then it can currently be in equilibrium in the sun even if almost all of the annihilation
cross-section at freeze-out was due to $p$-wave interactions (suppressed at current times), with only
a negligible amount due to $s$-wave interactions.

Similarly, for the case of long-range interactions, we plot in the bottom panel of fig.~\ref{Equilibrium} the
minimum $C$ required for the sun to be in equilibrium (assuming $\langle \sigma v \rangle = 1~\pb$).

\section{An Application to Neutrino Detectors}

We now consider an application of these tools to a specific detector.  We will focus on the
case of long-range interactions, because neutrino detectors are expected to have a major
advantage over direct detection experiments in this instance.  For both direct detection experiments
and neutrino searches, the measured event rate will
be proportional to the dark matter-nucleus scattering cross-section.
For the case of long-range interactions, the integrated dark matter-nucleus
scattering cross-section is roughly proportional to $C m_A^{-2} E_{min}^{-1} E_{max}^{-1}$, where $m_A$ is
the mass of the nucleus and $E_{min}$ is the minimum nuclear recoil energy which one can
measure.  For a direct detection experiment, $E_{min}$ is the recoil energy
threshold of the experiment, and is typically of order $2-10~\kev$.  For germanium-based
experiments (such as CDMS and CoGeNT), $m_A \sim 72~m_p$, while for xenon-based experiments
$m_A \sim 130~m_p$.  For neutrino searches, $E_{min}$ is the minimum recoil energy such that
dark matter is captured and can annihilate in the core of the sun.  We thus find
$E_{min} \approx (1/2) m_X u^2 \sim 2-5~\kev$,
for dark matter in the mass range considered here.  However, $m_A$ is the mass of the target
nucleus in the sun and is very small for some elements that contribute significantly to capture in the sun.
For example, hydrogen contributes $\sim 3\%$ of the dark matter capture rate for low-mass dark matter, and $m_H = m_p$.
So one can expect the sensitivity of a neutrino search for low-mass dark matter
with long range interactions to be significantly enhanced ($\sim 10^2 -10^3$) compared to direct detection experiments.

We compare the sensitivity of liquid scintillation (LS) neutrino detectors to that
of CDMS.  It was shown in~\cite{Learned:2009rv} that liquid scintillation neutrino-detectors
can determine the flavor and direction of leptons produced by a charged-current interaction using the
timing of the first photons which reach the photomultiplier tubes.
We will focus on a search for electron neutrinos producing
fully-contained electron/positron events.  An advantage of this strategy is that the atmospheric
electron neutrino background is significantly smaller than that of mu neutrinos.  It was estimated that
liquid scintillation neutrino detectors can provide almost absolute lepton flavor
discrimination, and electrons of the energy range we consider can be measured
with an angular resolution $\alt 1^\circ$.  It was also estimated that the neutrino energy could be
determined (from the energy and direction of the produced charged lepton, as well as total energy
deposition) with a resolution $\sim 1-3\%$.

We will consider the sensitivity of a LS neutrino detector with a spherical fiducial volume $V_0 \sim 1000~\m^3$
and 2135 live-days of data (these are roughly the specifications of KamLAND).
We estimate the neutrino detector's sensitivity utilizing the procedure outlined in section II.
The density of the liquid scintillator is taken to be $80\%$ that of water.
Following~\cite{Kumar:2011hi},
we define as ``fully-contained" an electron/positron event starting within the detector with
at least 10 radiation lengths
($\sim 4.3~\m$) contained within the detector.  Furthermore, the lepton event must
point back to the sun within a half-angle $\theta_{cone} = 20^\circ \sqrt{10~\gev / E_\nu}$,
and the energy of the neutrino must be obey $E_\nu \geq 1.5~\gev$.
We then find~\cite{Kumar:2011hi}
\bea
\int dV\,  \eta (r) \times \epsilon (r,z) &\sim& \eta \times {2\over 3} \times {1\over 2} V_0 ,
\eea
where the factor $2/3$ is the fraction of lepton events which will be within $\theta_{cone}$,
and the factor $1/2$ is the fraction of the fiducial volume which will yield
fully-contained events.
It was
shown in~\cite{Kumar:2011hi} that, using the background estimate of~\cite{Honda:2011nf}, one would expect
less than 5 electron/positron events satisfying these cuts arising from atmospheric neutrinos during the
specified runtime.
We will thus consider a model which would produce 10 signal events arising from dark matter annihilating
in the sun as being excludable.

To determine the electron (anti-)neutrino flux at Kamioka, we have used the WimpEvent routine, run
on the Hawaii Open Supercomputing Center cluster.  Of the $10^7$ dark matter annihilations which were simulated
(as described in section III) for each annihilation channel and value of $m_X$, $2\times 10^6$ were used to
compute the neutrino spectra at the detector.  The effect of neutrino propagation through the earth
typically suppresses $\langle Nz \rangle$ by $\sim 25-50\%$ (depending on the mass and
the annihilation channel).

For CDMS, we will roughly estimate the sensitivity to dark matter with
long-range interactions from their published bounds~\cite{Ahmed:2010wy} on dark matter with
elastic contact isospin-invariant interactions.
The bound CDMS can place on $C$ can be related to its bounds on $\sigmaSI^p$ by the relation
\bea
C^{bound} &=& \sigmaSI^{p(bound)} {m_{Ge}^2 \over 4\pi \alpha^2 \mu_p^2}
{\int_{u_{min}}^{\infty} du\, [f(u)/u] w^2 E_{max}^{-1} \int_{E_{thr}}^{E_{max}} dE_R\, |F_{Ge} (E_R)|^2
\over
\int_{u_{min}}^{\infty} du\, [f(u)/u] w^2 E_{max}^{-1} \int_{E_{thr}}^{E_{max}} dE_R\, |F_{Ge} (E_R)|^2 / E_R^2} ,
\eea
where $u$ is the velocity of a dark matter particle far from the sun, and $w$ is the
is the velocity of the same particle once it has reached the surface of the earth.  When it reaches the
surface of the earth, the kinetic energy of the particle has increased by an amount equal to the
change in the gravitational potential energy.  The change in the gravitational potential energy due to the
sun and the earth ($V_{sun}$ and $V_{earth}$, respectively) can be written as $\Delta V_{sun,earth} = -(1/2) m_X v_{sun,earth}^2$, where
$v_{sun} \approx 42.1~\km / \s$ is the escape velocity
of the sun at the radius of the earth's orbit, and $v_{earth} \approx 11.2~\km / \s$ is the escape velocity
of the earth at the surface of the earth.  Using the relation $\Delta E_{kinetic} = -\Delta V_{sun} - \Delta V_{earth}$,
we find $w = (u^2 + v_{sun}^2 + v_{earth}^2)^{1\over 2}$.

The maximum recoil energy which can be
transferred to a germanium nucleus is $E_{max}=2 m_X^2 m_{Ge}  w^2 / (m_X+m_{Ge})^2$.
We assume a threshold energy  $E_{thr} = 2~\kev$~\cite{Ahmed:2010wy}.
$u_{min}$ is the minimum dark matter velocity (far from the sun) such that
scattering with $E_R > E_{thr}$ is kinematically possible, and is given by the expression
$u_{min}^2 = \max [(m_{Ge} E_{thr}/ 2m_r^2 )-v_{sun}^2-v_{earth}^2,0]$.

We again assume a Gaussian form factor $F_A (E_R)$; for the recoil energy range of interest, the
Gaussian form factor for germanium differs from the Helm form factor~\cite{Helm} by at most $6\%$.
We will assume a Maxwell-Boltzmann velocity distribution with $\bar v = 270~\km / \s$, and
that the galactic escape velocity is $600~\km / \s$.
We will also assume a constant
efficiency for events with recoil energy greater than the threshold energy to appear in the
CDMS low-energy analysis band.
Due to this assumption, the result
shown here should be regarded as only an estimate of the sensitivity CDMS could obtain
with present data to dark matter models with long-range interactions.

\begin{figure}[tb]
\includegraphics[width=0.95\columnwidth]{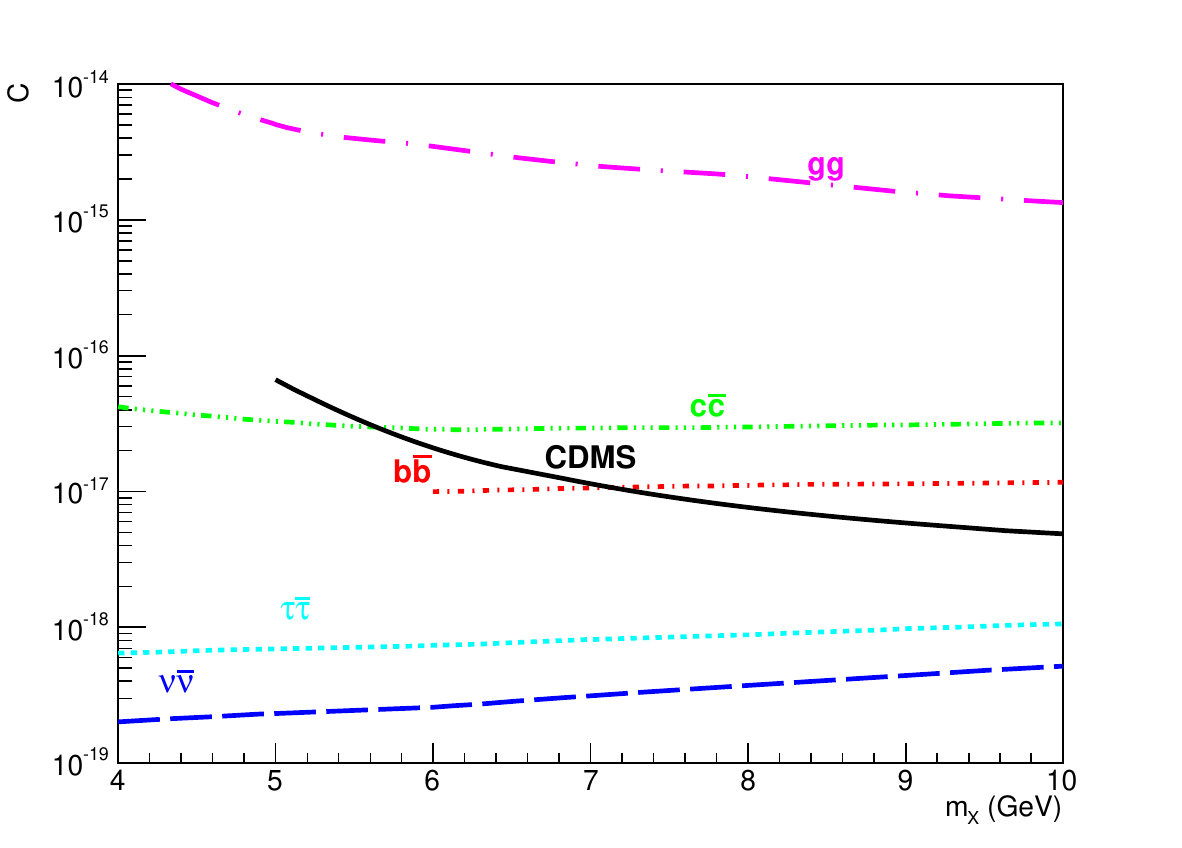}
\vspace*{-.1in}
\caption{Sensitivity to $C$  of CDMS and a 1 kT LS detector (2135 live days of data) for
low-mass dark matter with isospin-invariant elastic long-range interactions.  LS detector
sensitivity is shown assuming annihilation to either the $\tau$, $b$, $c$, $g$ and $\nu$
(flavor-independent) channels. }
\label{Sensitivity}
\end{figure}

The estimated sensitivity of CDMS and a 1 kT liquid scintillation neutrino detector
are plotted in figure~\ref{Sensitivity}.  For the LS neutrino detector, we assume 2135 live-days of
data, and assume that dark matter annihilates exclusively to either $\tau \bar \tau$,
$b \bar b$, $c \bar c$, $gg$ or $\nu \bar \nu$ (with equal coupling to all three neutrino flavors).
In~\cite{Kumar:2011hi}, it was shown that the sensitivity of CDMS to $10~\gev$ dark matter
with isospin-invariant elastic contact interactions is roughly an order of magnitude greater
than that of a 1 kT LS detector.  Our calculation of the relative sensitivities of CDMS and
a 1 kT LS detector to dark matter with long-range interactions bears out our original estimate of
a roughly $10^2-10^3$ relative enhancement in sensitivity for the LS detector.  Note that, for the models to which
KamLAND would be sensitive, the sun would be in equilibrium (see figure~\ref{Equilibrium}) even if the annihilation cross-section
were significantly smaller than $1~\pb$ (assuming standard astrophysical assumptions).  If
the sun is not in equilibrium as a result
of deviations from these assumptions, then the constraints which would
be possible from neutrino detectors would be significantly suppressed.

We will not attempt a quantitative estimate of the sensitivity of XENON100 to dark matter with
long-range interactions.  XENON100's recoil energy threshold is defined in terms of scintillation
photoelectrons; the detector's scintillation response to recoil energy (${\cal L}_{eff}$) is not
measured for low recoil energies.  Moreover, bounds from XENON100 are generated assuming that
the number of photoelectrons is determined by a Poisson distribution.  Even some low energy recoils can
thus produce enough scintillation photoelectrons to exceed the threshold.
As a result of the uncertainties in the detector response at low recoil, an attempt to estimate
the event rate expected at XENON100 for
dark matter with long-range interactions is beyond the scope of this work.
We will simply note that the recoil energy range for which XENON100 is sensitive is at best
comparable to that of CDMS, while a xenon nucleus is roughly twice as heavy as that of germanium.
This implies that the sensitivity of XENON100 relative to CDMS will be suppressed by roughly a factor
of 4 for the case of long-range interactions.

Recent hints of low-mass dark matter have potentially been seen by the DAMA~\cite{Bernabei:2010mq},
CoGeNT~\cite{Aalseth:2011wp} and CRESST~\cite{Angloher:2011uu}
experiments.  Dark matter models with long-range interactions have been discussed as a possible
way of reconciling the data from these experiments with the constraints from other direct
detection experiments~\cite{Fornengo:2011sz,Foot:2012rk}.  Long-range interactions can affect not only the magnitude
of the overall excesses seen by CoGeNT and CRESST, but can also affect the modulation seen by
DAMA and CoGeNT.  Although models with long-range interactions can provide a better fit to the
overall excesses, they sometimes provide a worse fit to the modulation signals.
We will not attempt to define a region of parameter-space
for dark matter with long-range interactions which could match the DAMA, CoGeNT or CRESST data.  This would require
a detailed matching of the expected event spectrum with that observed by the experiments, which is
beyond the scope of this work (and perhaps premature, given the issues raised in~\cite{CollarTaup2011}).
However, since CoGeNT also uses germanium as the target material, one would expect a 1 kT LS detector
to be easily sensitive to dark matter models with long-range interactions which could potentially explain the
data of CoGeNT.  Moreover, one should note that the low-mass CRESST region is consistent with scattering
from both oxygen and calcium.  Given the difference in mass, one would expect long-range interactions with
calcium to be suppressed by roughly a factor 4 relative to oxygen, as compared to the case of contact
interactions.

Finally, we can consider the sensitivity of LBNE (Long-Baseline Neutrino Experiment).
We will assume the detector target material is liquid argon (configuration 2)~\cite{Akiri:2011dv},
with a total fiducial volume of roughly 51 kT.
Liquid argon-based neutrino detectors are expected to have very good event reconstruction; we will
assume that liquid argon detectors permit a reconstruction of charged lepton flavor, energy and direction with
at least the same resolution as liquid scintillation detectors.
We then find
\bea
\int dV\,  \eta (r) \times \epsilon (r,z) &\sim&  {2\over 3} \times (3.1 \times 10^{34})
\eea
where the factor $2/3$ again arises from the fraction of charged lepton events which would
point back to the sun within angle $\theta_{cone}$.  For a detector as large as LBNE, almost
the entire fiducial volume can produce fully-contained events.
We then see that the sensitivity which KamLAND could obtain with its 2135 day data set
could be obtained by LBNE with only $\sim 17$ days of data.

Note that the possibility of dominant annihilation to leptonic channels is not inconsistent with
dark matter-nucleus scattering which is large enough to be probed by neutrino detectors.  For example,
it could be that dark matter-quark scattering is mediated by an effective operator which permits
velocity-independent, spin-independent scattering, but does not permit $s-$wave annihilation (an
example of such an operator is $\bar X X \bar q q$).  In this case, the dark matter-nucleus scattering
cross-section could be reasonably large, while the the cross-section for dark matter to annihilate to quarks
would be $v^2$-suppressed.  If dark matter coupled to leptons through an operator which permitted
$s-$wave annihilation (an example of such an operator would be $ \bar X \gamma_\mu X \bar f \gamma^\mu f$,
if the dark matter were a Dirac fermion),
then the dark matter would mostly annihilate to leptons.

\section{Conclusions}

We have computed the capture rates and neutrino spectra which are relevant for
neutrino-based searches for low-mass dark matter in the sun.  The neutrino
spectra are presented at a distance of 1 AU from the sun, accounting for matter
effects in the sun, and vacuum oscillations (assuming a normal hierarchy and $\theta_{13}=10^\circ$).
The capture rates
have been found assuming either elastic contact, elastic long-range, or inelastic
contact interactions.  These are the tools required for a neutrino detector to search
for dark matter annihilating in the sun.

As an application of these tools, we plot the sensitivity of a 1 kT LS detector,
with 2135 days of data, to low-mass dark matter with isospin-invariant elastic
long-range interactions with Standard Model nucleons.  We have found that neutrino
detectors have a greatly enhanced sensitivity to dark matter with long-range interactions,
relative to leading direct detection experiments such as CDMS.  This
enhancement is readily understood; in the case of long-range interactions, the scattering
matrix element is inversely proportional to $q^2 = 2 m_A E_R$.  Scattering rates in detectors
with heavy targets, such as germanium and xenon, are heavily suppressed.  Dark matter capture in
the sun involves scattering from low-mass targets such as hydrogen and helium, implying
that these scattering rates will see a relative enhancement.  A LS neutrino detector with the exposure
already available to KamLAND could have a sensitivity up to 2 orders of magnitude greater
than that of CDMS.  LBNE (with a 51 kT liquid argon target) could achieve similar sensitivity
with roughly 17 days of data.

We have also found that low-mass dark matter with inelastic contact interactions can be probed by
neutrino detectors even for $\delta m_X \sim 50~\kev$.  This implies that neutrino detectors can be
sensitive to inelastic dark matter models which are more difficult to probe on earth, because
gravitational infall allows inelastic scattering in the sun for models where inelastic scattering
would not be kinematically possible on earth.

The choice $\theta_{13}=10^\circ $ is consistent with
recent data from the Daya Bay experiment~\cite{An:2012eh}.  The neutrino spectrum is slightly different
from the $\theta_{13}= 0^\circ $ case, with the difference most noticeable in the case of annihilation entirely
to neutrinos.  For searches involving upward-going leptons, there will also
be a modification to the neutrino spectrum due to passage through the earth.  This effect will depend
on the location of the detector; for any particular detector, one can obtain the appropriate neutrino spectra
by running the WimpEvent program, inputting the data files for the neutrino spectrum at 1 AU found at
\url{http://www.phys.hawaii.edu/~superk/post/spectrum}.

It is worth noting that a direct detection experiment with a target molecule containing
hydrogen would also be expected to have enhanced sensitivity to dark matter with long-range
interactions.  Gaseous time projection chambers (such as DRIFT~\cite{Daw:2011wq},
DMTPC~\cite{Monroe:2011er}, $D^3$~\cite{Vahsen:2011qx},
MIMAC~\cite{Santos:2011kf} and NEWAGE~\cite{Miuchi:2010hn})
using hydrocarbon targets may be
well-suited for this type of search.

Specific dark matter models with long-range interactions may have solar capture rates that are
enhanced by collective effects, such as multiple scattering.  Neutrino searches thus have
enhanced sensitivity to such models, and current data may already provide tight constraints.
It would be interesting to consider such models in more detail.

\section{Acknowledgments}

We gratefully acknowledge K.~Choi, D.~Marfatia, M.~Sakai, P.~Subramoney and
S.~Vahsen for useful discussions.  We also thank the Hawaii Open Supercomputing
Center.  This work is supported in
part by the Department of Energy under Grant~DE-FG02-04ER41291.

\appendix

\section{Neutrino Spectra}

In this appendix, we plot the neutrino and anti-neutrino
spectra at a distance 1 AU from the sun, assuming dark matter
annihilation exclusively to either the $b \bar b$ or
$\tau \bar \tau$ channels.  Each spectrum was generated by
simulating $10^7$ annihilations using the method described in the text.
Here, $z\equiv E_\nu / m_X$, and neutrino oscillations are generated assuming
$\theta_{13}=10^\circ$.

\newpage

\begin{figure}[tb]
\includegraphics[width=1.0\columnwidth]{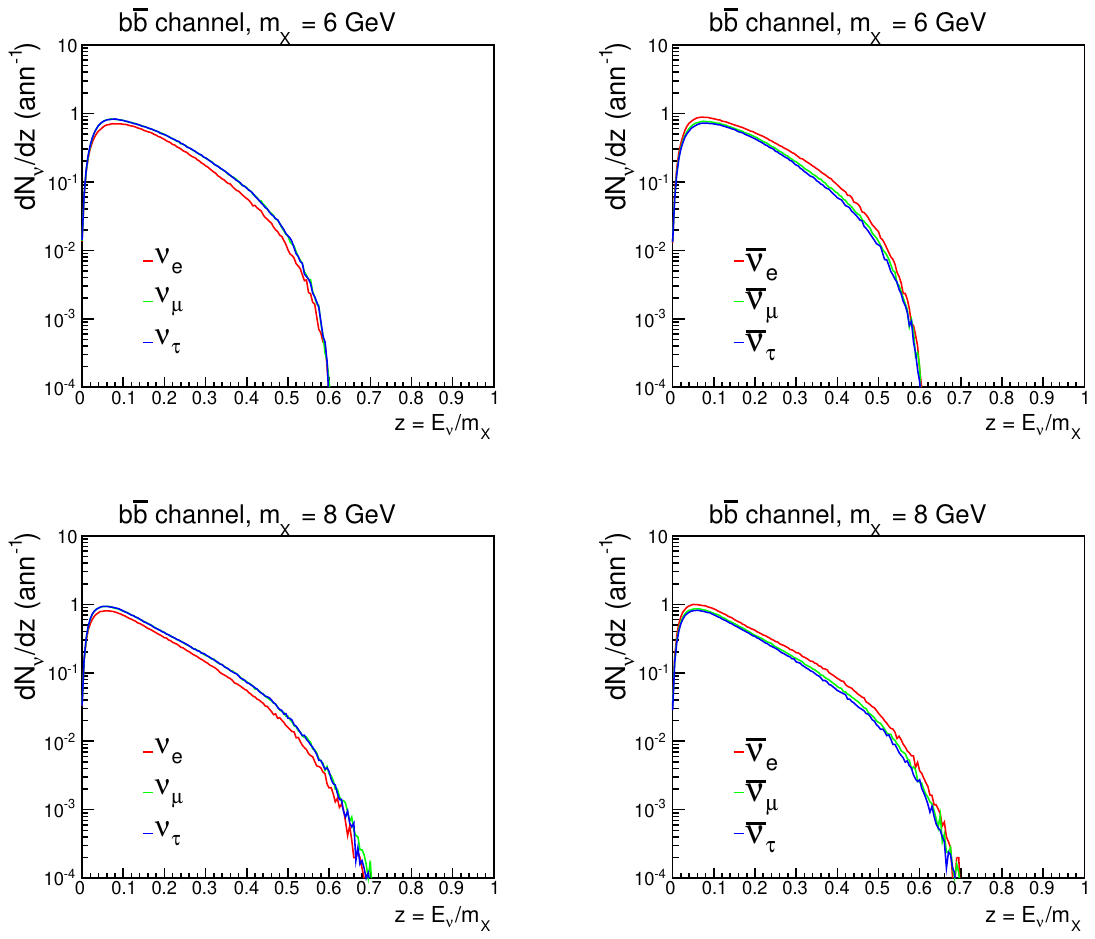}
\vspace*{-.1in}
\caption{Neutrino spectra (left panels) and anti-neutrino spectra (right panels)
at 1 AU for dark matter annihilation to the $b \bar b$ channel.  The spectra for
$\nu_e$($\bar \nu_e$), $\nu_\mu$($\bar \nu_\mu$), and $\nu_\tau$($\bar \nu_\tau$)
are shown in red, green, and blue, respectively.  Spectra are shown for $m_X = 6,8~\gev$.}
\end{figure}

\newpage

\begin{figure}[tb]
\includegraphics[width=1.0\columnwidth]{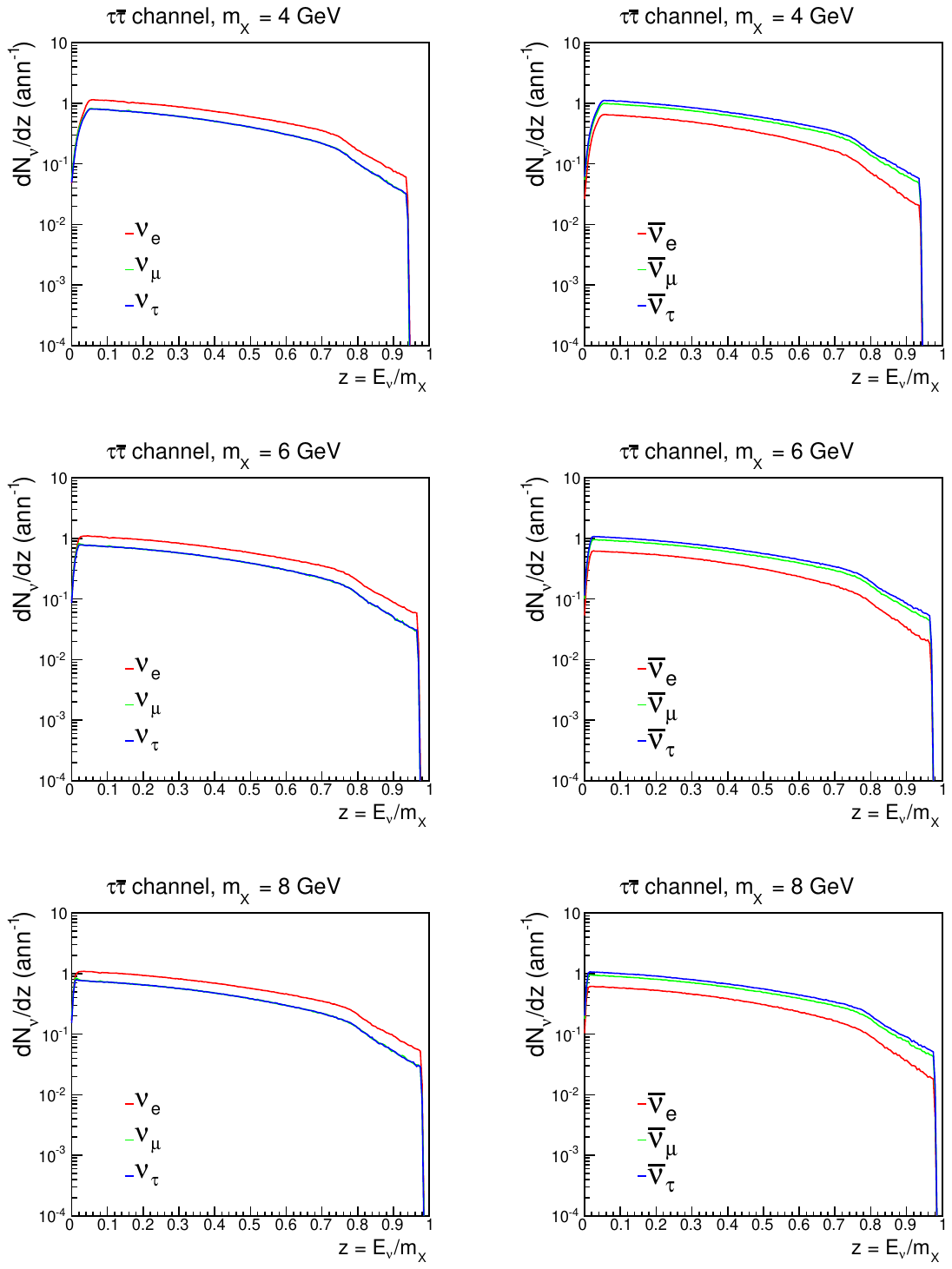}
\vspace*{-.1in}
\caption{Neutrino spectra (left panels) and anti-neutrino spectra (right panels)
at 1 AU for dark matter annihilation to the $\tau \bar \tau$ channel.  The spectra for
$\nu_e$($\bar \nu_e$), $\nu_\mu$($\bar \nu_\mu$), and $\nu_\tau$($\bar \nu_\tau$)
are shown in red, green, and blue, respectively.  Spectra are shown for $m_X = 4,6,8~\gev$.}
\end{figure}

\end{document}